\documentclass[twocolumn,superscriptaddress,amsmath,amssymb,floats]{revtex4-1}
\usepackage[latin9]{inputenc}
\usepackage{amsmath}
\usepackage{graphicx}
\usepackage{dsfont}

\makeatletter

\newcommand*\LyXThinSpace{\,\hspace{0pt}}

\@ifundefined{textcolor}{}
{%
 \definecolor{BLACK}{gray}{0}
 \definecolor{WHITE}{gray}{1}
 \definecolor{RED}{rgb}{1,0,0}
 \definecolor{GREEN}{rgb}{0,1,0}
 \definecolor{BLUE}{rgb}{0,0,1}
 \definecolor{CYAN}{cmyk}{1,0,0,0}
 \definecolor{MAGENTA}{cmyk}{0,1,0,0}
 \definecolor{YELLOW}{cmyk}{0,0,1,0}
}

\batchmode

\def\input@path{{\string"/Users/rafaelmf/Google Drive/Morten Notes Shared/Nematics/Paper draft/\string"/}}

\@ifundefined{textcolor}{}{%
 \definecolor{BLACK}{gray}{0}
 \definecolor{WHITE}{gray}{1}
 \definecolor{RED}{rgb}{1,0,0}
 \definecolor{GREEN}{rgb}{0,1,0}
 \definecolor{BLUE}{rgb}{0,0,1}
 \definecolor{CYAN}{cmyk}{1,0,0,0}
 \definecolor{MAGENTA}{cmyk}{0,1,0,0}
 \definecolor{YELLOW}{cmyk}{0,0,1,0}
}

\usepackage{epstopdf}
\usepackage{color}
\usepackage{array}
\usepackage{dcolumn}
\usepackage{bm}

\newcolumntype{L}[1]{>{\raggedright\let\newline\\\arraybackslash\hspace{0pt}}m{#1}}
\newcolumntype{C}[1]{>{\centering\let\newline\\\arraybackslash\hspace{0pt}}m{#1}}
\newcolumntype{R}[1]{>{\raggedleft\let\newline\\\arraybackslash\hspace{0pt}}m{#1}}

\graphicspath{{Impurities and Gaps/Plots/}}

\newcommand{\mbf}[1]{\mathbf{#1}}

\usepackage{tikz}
\usetikzlibrary{calc,intersections}
\usetikzlibrary{shadings}
\usepgflibrary{shadings}

\makeatother

\begin{document}

\title{Spin-Driven Nematic Instability of the Multi-Orbital Hubbard Model:
\\
 Application to Iron-Based Superconductors}

\author{Morten H. Christensen}

\affiliation{School of Physics and Astronomy, University of Minnesota, Minneapolis,
MN 55455, USA}

\affiliation{Niels Bohr Institute, University of Copenhagen, Juliane Maries Vej
30, DK-2100, Denmark}

\author{Jian Kang}

\affiliation{School of Physics and Astronomy, University of Minnesota, Minneapolis,
MN 55455, USA}

\author{Brian M. Andersen}

\affiliation{Niels Bohr Institute, University of Copenhagen, Juliane Maries Vej
30, DK-2100, Denmark}

\author{Rafael M. Fernandes}

\affiliation{School of Physics and Astronomy, University of Minnesota, Minneapolis,
MN 55455, USA}
\begin{abstract}
Nematic order resulting from the partial melting of density-waves
has been proposed as the mechanism to explain nematicity in iron-based
superconductors. An outstanding question, however, is whether the
microscopic electronic model for these systems -- the multi-orbital
Hubbard model -- displays such an ordered state as its leading instability.
In contrast to usual electronic instabilities, such as magnetic and
charge order, this fluctuation-driven phenomenon cannot be captured
by the standard RPA method. Here, by including fluctuations beyond
RPA in the multi-orbital Hubbard model, we derive its nematic susceptibility
and contrast it with its ferro-orbital order susceptibility, showing
that its leading instability is the spin-driven nematic phase. Our
results also demonstrate the primary role played by the $d_{xy}$
orbital in driving the nematic transition, and reveal that high-energy
magnetic fluctuations are essential to stabilize nematic order in
the absence of magnetic order. 
\end{abstract}
\maketitle
The elucidation of electronic Ising-nematic order \cite{chandra01}
-- the state in which electronic degrees of freedom spontaneously
lower the point-group symmetry of the system -- has become an important
problem in unconventional superconductors \cite{fernandes02,kivelson02}.
In both pnictides \cite{fisher01,Davis10,ZXshen11,Matsuda12,fisher02,degiorgi01}
and cuprates \cite{Keimer08,daou01,Lawler10}, the experimentally
observed nematic order has been proposed to arise from the partial
melting of an underlying spin density-wave (SDW) \cite{Kivelson08,Sachdev08,fernandes01,Abrahams_Si}
or charge density-wave (CDW) \cite{Kivelson98,Kivelson_PNAS14,Chubukov14}
stripe-order. This mechanism is based on robust symmetry considerations.
Consider for concreteness the stripe SDW case: the ground state has
an $O\left(3\right)\times Z_{2}$ degeneracy, with $O(3)$ denoting
the direction of the magnetic order parameter in spin space, and $Z_{2}$
denoting the selection of the SDW ordering vector $\mathbf{Q}_{X}=\left(\pi,0\right)$
or $\mathbf{Q}_{Y}=\left(0,\pi\right)$ (in the CDW case, the system
has an $O\left(2\right)\times Z_{2}$ degeneracy). Fluctuations in
layered systems suppress the continuous ($O(3)$ or $O(2)$) and the
discrete ($Z_{2}$) symmetries differently, favoring an intermediate
regime in which only the $Z_{2}$ symmetry is broken \cite{fernandes01}.
Because the $Z_{2}$ symmetry distinguishes between two ordering vectors
related by a $90^{\circ}$ rotation, its breaking implies a tetragonal-to-orthorhombic
transition, and therefore nematic order.

Although this mechanism for spin-driven (or charge-driven) nematic
order has been established in simplified low-energy models for pnictides
\cite{Kivelson08,Sachdev08,eremin01,fernandes01,fanfarillo01} and
cuprates \cite{Kivelson_PNAS14,Chubukov14}, it remains hotly debated
whether more realistic microscopic models display nematic order as
the leading electronic instability. 
the cuprates, a sensible microscopic model is the single-band Hubbard
model, whose phase diagram has been reported to display nematic correlations
in the strong-coupling regime~\cite{okamoto01,su01}. For the pnictides,
due to the $3d^{6}$ configuration of Fe and to the small crystal
field splittings, a five-orbital Hubbard model, including Hund's rule
interactions, is a more appropriate starting point~\cite{Kuroki08,kemper01}.
Furthermore, because many pnictides display metallic behavior, a weak-coupling
analysis of this intricate model can reveal important information
about the underlying physics of these materials. Indeed, conventional
RPA approaches have been employed to study the onset of SDW, CDW,
and ferromagnetism. However, in contrast to these usual
electronic instabilities, the standard RPA approach does not capture
the nematic instability even qualitatively, as we show below, making
it difficult to assess whether the realistic multi-orbital Hubbard
model has a tendency towards nematic order.

In this Letter, we extend the standard RPA approach and derive the
nematic susceptibility of an arbitrary multi-orbital Hubbard model.
The fluctuations included in this formalism arise solely from the
non-interacting part of the Hamiltonian, such that interactions are
treated at the same order as in the typical RPA method. We apply
this formalism to the case of SDW-driven nematicity in iron pnictides,
and establish that the leading instability of the five-orbital interacting
model is a spin-driven nematic phase for a wide range of parameters.
In general, we find that nematic order exists in a narrow $T$ range
above the magnetic transition line, in agreement with experiments
in the pnictides \cite{reviews}. However, magnetic fluctuations at
higher energies can induce a sizable splitting between the two transitions,
particularly in the regime where the SDW transition is suppressed
to zero. We propose that this effect may be relevant to understanding
the unusual nematic phase of FeSe \cite{Chubukov_Fernandes15,Mazin15,Kivelson_Lee_15,Si15,Kontani15}.
Previously, the investigation of the multi-orbital Hubbard model in
Ref.~\cite{fanfarillo01} revealed the importance of the orbital
content of the Fermi surface in the low-energy spin-nematic model
of the pnictides. Here, we find from the orbitally-resolved nematic
susceptibility that whereas the $d_{xz}$, $d_{yz}$, and $d_{xy}$
orbitals contribute almost equally to the SDW instability, the $d_{xy}$
orbital plays a stronger role in driving the nematic instability.
Finally, we compare the nematic susceptibility with the RPA-derived
ferro-orbital order susceptibility. 
work provides a promising route to search for nematicity in different
compounds, as it is compatible with \emph{ab initio} approaches and
also with methods that include the effects of moderate interactions,
such as LDA+DMFT \cite{park01,Medici13}.

Our starting point is the multi-orbital Hubbard model with onsite
interactions \cite{kemper01,gastiasoro01}. The non-interacting part
is given by $\mathcal{H}_{0}=\sum_{\mu,\nu}\left(\epsilon_{\mu\nu}(\mbf{k})-\tilde{\epsilon}\delta_{\mu\nu}\right)c_{\mbf{k}\mu\sigma}^{\dagger}c_{\mbf{k}\nu\sigma}$,
where $c_{\mbf{k}\mu\sigma}^{\dagger}$ creates an electron with momentum
$\mathbf{k}$ and spin $\sigma$ at orbital $\mu=1,...,N_{\mathrm{orb}}$
and the hopping parameters $\epsilon_{\mu\nu}(\mbf{k})$ are determined
from tight-binding fits to \emph{ab initio} calculations (sums over
spin and momentum indices are left implicit). The four onsite interaction
terms correspond to the intra-orbital Hubbard term, $\mathcal{H}_{U}=U\sum_{\mu}n_{\mbf{q}\mu\uparrow}n_{\mbf{-q}\mu\downarrow}$,
the inter-orbital Hubbard term, $\mathcal{H}_{U'}=U'\sum_{\mu<\nu}n_{\mbf{q}\mu\sigma}n_{\mbf{-q}\nu\sigma'}$,
the Hund's rule coupling, $\mathcal{H}_{J}=J\sum_{\mu<\nu}c_{\mbf{k+q}\mu\sigma}^{\dagger}c_{\mbf{k}\nu\sigma}c_{\mbf{k'-q}\nu\sigma'}^{\dagger}c_{\mbf{k'}\mu\sigma'}$,
and the pair-hopping term $\mathcal{H}_{J'}=J'\sum_{\mu<\nu}c_{\mbf{k+q}\mu\sigma}^{\dagger}c_{\mbf{k'-q}\mu\bar{\sigma}}^{\dagger}c_{\mbf{k'}\nu\bar{\sigma}}c_{\mbf{k}\nu\sigma}$.
These coefficients are related by $U'=U-2J$ and $J'=J$. Previous
approaches considered the nematic susceptibility of a spin-fermion
model \cite{Dagotto13}; here, we will focus on the Hubbard model
within RPA. The mechanism in which nematic order arises from the partial
melting of an SDW or a CDW requires fluctuations at two momenta related
by $90^{\circ}$, in general $\mathbf{Q}_{1}=\left(\frac{\pi}{n},0\right)$
and $\mathbf{Q}_{2}=\left(0,\frac{\pi}{n}\right)$, with integer $n$.
Although our formalism can be extended in a straightforward way to
arbitrary $n$, hereafter we focus on $n=1$. 
to make contact with the pnictides, we consider the SDW channel. Performing
a Hartree-Fock decoupling of $\mathcal{H}$ in both the $\mbf{q}=0$
charge channel and the $\mbf{q}=\mbf{Q}_{i}$ SDW channel: 
\begin{eqnarray}
\mathcal{H}^{\text{MF}} & = & \sum_{\substack{\mbf{k}}
}\left(\epsilon_{\mu\nu}(\mbf{k})-\tilde{\epsilon}_{\nu}\delta_{\mu\nu}\right)c_{\mbf{k}\mu\sigma}^{\dagger}c_{\mbf{k}\nu\sigma}\nonumber \\
 & - & \frac{1}{2}\sum_{\substack{\mbf{kq}}
}\mbf{M}_{\mbf{q}\,\mu}^{i}\cdot c_{\mbf{k-q+Q}_{i}\mu\sigma}^{\dagger}\boldsymbol{\sigma}_{\sigma\sigma'}c_{\mbf{k}\mu\sigma'}\,,\label{eq:hamiltonian}
\end{eqnarray}
where $\tilde{\epsilon}_{\nu}$ incorporates the changes in the mean-field
densities and $\mbf{M}_{\mbf{q}\,\mu}^{i}=\frac{1}{2}\sum_{\mbf{k}}U_{\mu}^{\rho}\langle c_{\mbf{k}+\mbf{q}+\mbf{Q}_{i}\rho\sigma}^{\dagger}\boldsymbol{\sigma}_{\sigma\sigma'}c_{\mbf{k}\rho\sigma'}\rangle$
are the SDW order parameters with $i=X,Y$. The interaction matrix
$U_{\mu}^{\rho}$ is $U_{a}^{a}=U$ and $U_{b\neq a}^{a}=J$. We consider
only intra-orbital magnetism, since previous Hartree-Fock calculations
revealed that in the ground state the intra-orbital SDW order parameters
are the dominant ones \cite{gastiasoro01}. In the standard RPA approach
for the SDW instability, the electronic degrees of freedom are integrated
out, yielding the quadratic magnetic free energy:

\begin{eqnarray}
F_{\text{mag}}^{(2)}[\mbf{M}_{\mu}^{i}] & = & \sum_{q,i=X,Y}\left[\chi_{i}^{\mu\nu}(q)\right]^{-1}\mbf{M}_{q,\mu}^{i}\cdot\mbf{M}_{-q,\nu}^{i}\,,\label{eq:second_order_free_energy}
\end{eqnarray}
with the magnetic propagator $\chi_{i}^{\mu\nu}(q)$ 
\begin{eqnarray}
\chi_{i}^{\mu\nu}(q) & = & \big[\left(U_{\nu}^{\mu}\right)^{-1}+\sum_{k}\mathcal{G}^{\nu\mu}(k)\mathcal{G}_{i}^{\mu\nu}(k+q)\big]^{-1}\,,\label{eq:mag_susc}
\end{eqnarray}
where $\mathcal{G}_{i}^{\mu\nu}(k)\equiv\mathcal{G}^{\mu\nu}(k+\mbf{Q}_{i})$
is the Green's function in orbital basis, $q=(\mbf{q},\Omega_{n})$,
$\sum_{q}=T/N_{\mbf{q}}\sum_{\mbf{q}}\sum_{\Omega_{n}}$, and $\Omega_{n}=2n\pi T$
is the Matsubara frequency. The RPA magnetic susceptibility $\left\langle \mbf{M}_{q,\mu}^{i}\cdot\mbf{M}_{-q,\nu}^{i}\right\rangle $
is proportional to and diverges at the same temperature as the magnetic
propagator $\chi_{i}^{\mu\nu}(q)$. Note that the tetragonal symmetry
of the system implies that a peak of $\chi_{i}^{\mu\nu}(q)$ at $\mathbf{Q}_{X}=\left(\pi,0\right)$
will be accompanied by an equal peak at $\mathbf{Q}_{Y}=\left(0,\pi\right)$.
Therefore, at this order in perturbation theory, the system does not
distinguish the case in which either $\mathbf{Q}_{X}$ or $\mathbf{Q}_{Y}$
is selected (single-\textbf{Q} order) from the case in which both
are selected (double-\textbf{Q} order), i.e. the standard RPA approach
is blind to nematicity. To remedy this problem, we go beyond the second-order
expansion of the free energy and calculate the quartic-order terms:
\begin{eqnarray}
F_{\text{mag}}^{(4)}[\mbf{M}_{\mu}^{X},\mbf{M}_{\mu}^{Y}] & = & \frac{1}{2}u^{\rho\nu\eta\mu}\left(\mbf{M}_{\rho}^{X}\cdot\mbf{M}_{\nu}^{X}+\mbf{M}_{\rho}^{Y}\cdot\mbf{M}_{\nu}^{Y}\right)\nonumber \\
 &  & \qquad\qquad\times\left(\mbf{M}_{\eta}^{X}\cdot\mbf{M}_{\mu}^{X}+\mbf{M}_{\eta}^{Y}\cdot\mbf{M}_{\mu}^{Y}\right)\nonumber \\
 & - & \frac{1}{2}g^{\rho\nu\eta\mu}\left(\mbf{M}_{\rho}^{X}\cdot\mbf{M}_{\nu}^{X}-\mbf{M}_{\rho}^{Y}\cdot\mbf{M}_{\nu}^{Y}\right)\nonumber \\
 &  & \qquad\qquad\times\left(\mbf{M}_{\eta}^{X}\cdot\mbf{M}_{\mu}^{X}-\mbf{M}_{\eta}^{Y}\cdot\mbf{M}_{\mu}^{Y}\right)\nonumber \\
 & + & 2w^{\rho\nu\eta\mu}\left(\mbf{M}_{\rho}^{X}\cdot\mbf{M}_{\nu}^{Y}\right)\left(\mbf{M}_{\eta}^{X}\cdot\mbf{M}_{\mu}^{Y}\right)\,,\label{eq:fourth_order_free_energy}
\end{eqnarray}

The quartic coefficients, whose expressions are shown explicitly in
Appendix \ref{app:fourth_order_coeff}, depend only
on the non-interacting Green's functions. Although interactions can
also contribute to them, as shown in Refs. \cite{XiaoyuWang15,Kontani12},
within the RPA approach these contributions are sub-leading and can
be neglected. The most relevant coefficient for the nematic instability
is $g^{\rho\nu\eta\mu}$, whose term distinguishes between single-\textbf{Q}
and double-\textbf{Q } order. Specifically, a Hubbard-Stratonovich
transformation of this term reveals the nematic order parameter $\langle\phi_{\mu\nu}\rangle\propto\left\langle \mbf{M}_{\mu}^{X}\mbf{M}_{\nu}^{X}\right\rangle -\left\langle \mbf{M}_{\mu}^{Y}\mbf{M}_{\nu}^{Y}\right\rangle $,
a rank-2 tensor in orbital space that breaks the tetragonal symmetry
of the system by making $X\neq Y$. The term with coefficient $u^{\rho\nu\eta\mu}$
is related to Gaussian magnetic fluctuations in both SDW channels,
while the term with coefficient $w^{\rho\nu\eta\mu}$ mainly distinguishes
between the two types of double-\textbf{Q }order \cite{XiaoyuWang15}.
Eq. (\ref{eq:fourth_order_free_energy}) is the multi-orbital generalization
of the magnetic free energy previously obtained in effective low-energy
models in the band basis, where the coefficient $g$ becomes a scalar
\cite{fernandes01}.

\begin{figure}
\centering \includegraphics[width=\columnwidth]{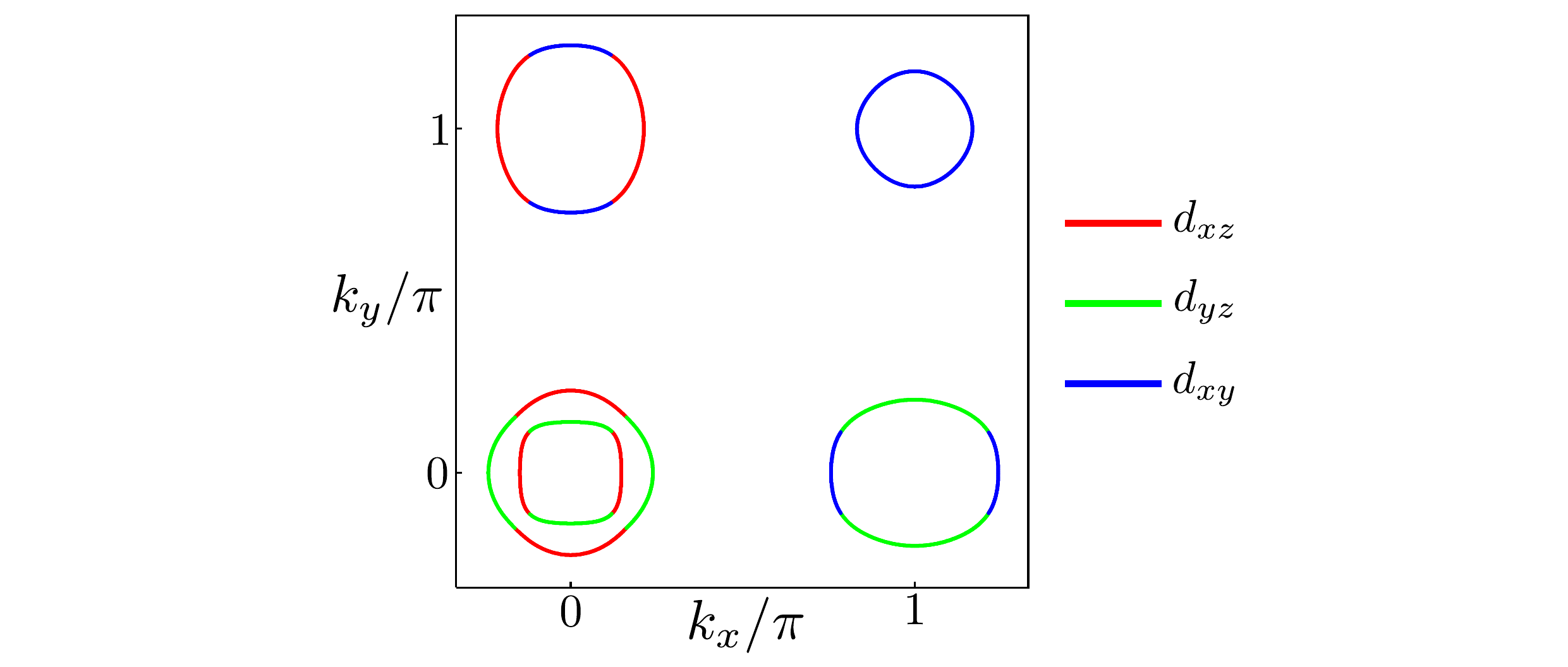}
\protect\protect\protect\protect\protect\protect\caption{\label{fig:fermi_surface} (Color online) Normal-state Fermi surface
based on the parameters of Ikeda \textit{et al}.~\cite{ikeda01}.
The colors indicate the dominant orbital contribution.}
\end{figure}

It is now possible to compute the static nematic susceptibility $\chi_{\text{nem}}^{\rho\nu\eta\mu}\propto\left\langle \phi_{\rho\nu}\phi_{\eta\mu}\right\rangle $
in the paramagnetic phase (see Appendix \ref{app:nem_susc} for details of the derivation): 
\begin{eqnarray}
\chi_{\text{nem}}^{\rho\nu\eta\mu} & = & \chi_{\text{nem},0}^{\eta\alpha\mu\beta}\left(\delta_{\rho\beta}\delta_{\nu\alpha}-g^{\rho\nu\gamma\delta}\chi_{\text{nem},0}^{\gamma\alpha\delta\beta}\right)^{-1},\label{eq:chi_nem}\\
\chi_{\text{nem},0}^{\rho\nu\eta\mu} & \equiv & \frac{1}{2}\sum_{q,i=X,Y}\chi_{i}^{\rho\nu}(q)\chi_{i}^{\eta\mu}(-q)\,.\label{eq:chi_bare_nem}
\end{eqnarray}
The orbitally-resolved nematic susceptibility $\chi_{\text{nem}}^{\rho\nu\eta\mu}$
is a rank-4 tensor that generalizes the scalar nematic susceptibility
derived previously for effective low-energy models \cite{fernandes03,Khodas15,Yamase15,Gallais15}.
The impact of the magnetic fluctuations encoded in the coefficient
$g^{\rho\nu\gamma\delta}$ is clear: if this term were absent, then
the (bare) nematic susceptibility would be merely a higher-order convolution
of the magnetic propagator, $\chi_{\text{nem},0}^{\rho\nu\eta\mu}$,
and therefore diverge at the same $T$ as the SDW susceptibility.
To establish whether the nematic susceptibility diverges already in
the paramagnetic phase, one needs to compute its leading eigenvalue
$\lambda^{(n)}$ from $\chi_{\text{nem}}^{\rho\nu\eta\mu}\Phi_{\rho\nu}^{(n)}=\lambda^{(n)}\Phi_{\eta\mu}^{(n)}$,
with $n=1,...,N_{\text{orb}}^{2}$. The structure of the corresponding
eigen-matrix $\Phi_{\eta\mu}^{(n)}$ reveals which orbitals promote
the nematic instability, and which orbitals favor a double-\textbf{Q
}structure with no underlying nematicity. We note that in principle
the Gaussian fluctuations associated with the term with coefficient
$u^{\rho\nu\eta\mu}$ can also renormalize the magnetic propagator
$\chi_{i}^{\rho\nu}$. However, because this effect merely renormalizes
the SDW transition temperature, we do not include it hereafter.

\begin{figure}
\centering \includegraphics[width=\columnwidth]{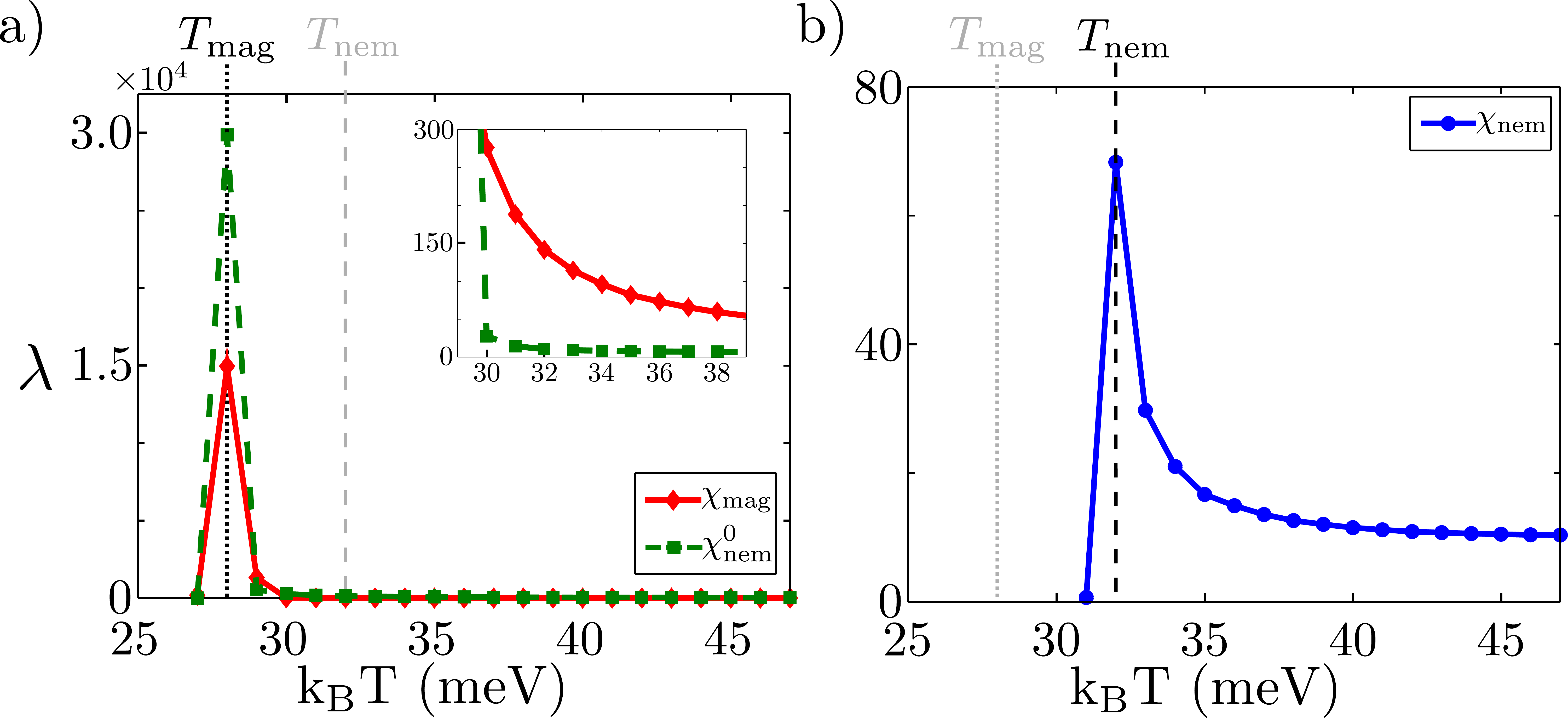}
\protect\protect\protect\protect\protect\protect\caption{\label{fig:nematic_and_magnetic_instabilities} (Color online) The
largest eigenvalues $\lambda$ of \textbf{(a)} the bare nematic susceptibility
$\chi_{\text{nem},0}^{\rho\nu\eta\mu}$, the $\mathbf{Q}_{X/Y}$ magnetic
propagator $\chi_{\text{mag}}$, and \textbf{(b)} the full nematic
susceptibility $\chi_{\text{nem}}^{\rho\nu\eta\mu}$ as a function
of $T$ for the case $n=6$. The inset in \textbf{(a)} shows the upturn
of the magnetic susceptibility as it diverges.}
\end{figure}

Equation~(\ref{eq:chi_nem}) is the RPA-generalized nematic susceptibility,
which can be compared on equal-footing with other RPA instabilities
of a weakly-interacting system described by a multi-orbital Hubbard
model. We apply this formalism to a five-orbital model for the iron-based
superconductors and contrast the nematic susceptibility to the ferro-orbital
RPA susceptibility. The hopping parameters are those from Ref. \cite{ikeda01},
whereas the interactions are set to $U=0.95$ eV and $J=U/4$ \cite{gastiasoro01}.
Small changes in these parameters do not alter our main results. The
Fermi surface for the occupation number $n=6$ is presented in Fig.
\ref{fig:fermi_surface}, consisting of three hole pockets at the
center and the corner of the Brillouin zone, and two electron pockets
at the borders of the Brillouin zone. 
that the $d_{xy}$ hole pocket at $\left(\pi,\pi\right)$ is not present
in all materials, as it depends on the Fe-As distance~\cite{Vildosola08,Calderon09}.

We evaluate Eqs.~(\ref{eq:mag_susc}) and (\ref{eq:chi_nem}) numerically
as functions of $T$ for various values of the occupation number $n$.
Consider first $n=6$: in Fig.~\ref{fig:nematic_and_magnetic_instabilities}(a),
we plot the $T$ dependence of the largest eigenvalue of the static
magnetic propagator $\chi_{i}^{\mu\nu}\left(0\right)$ as well as
the largest eigenvalue of the \emph{bare} nematic susceptibility $\chi_{\text{nem},0}^{\rho\nu\eta\mu}$.
Despite having different $T$ dependencies, both eigenvalues diverge
at the same temperature $T_{\mathrm{mag}}$, confirming our assertion
that the standard RPA is blind to the nematic instability. In Fig.~\ref{fig:nematic_and_magnetic_instabilities}(b),
we plot the largest eigenvalue of the \emph{full} nematic susceptibility
$\chi_{\text{nem}}^{\rho\nu\eta\mu}$, as given by Eq. (\ref{eq:chi_nem}).
Clearly, the eigenvalue diverges at $T>T_{\mathrm{mag}}$: this is
exactly the nematic transition temperature $T_{\mathrm{nem}}$.

Interestingly, our results reveal a relatively small splitting between
$T_{\mathrm{nem}}$ and $T_{\mathrm{mag}}$, with $T_{\mathrm{nem}}\approx1.14T_{\mathrm{mag}}$,
which resembles the small $T$-range in which a nematic-paramagnetic
phase is observed experimentally in the iron pnictides \cite{reviews}.
We caution, however, that this value should be understood as an upper
boundary for the splitting between the nematic and the actual magnetic
transition, since $\widetilde{T}_{\mathrm{mag}}$ calculated inside
the nematic state is generally larger than $T_{\mathrm{mag}}$ calculated
in the tetragonal state. Furthermore, the value for $T_{\mathrm{mag}}$
obtained via RPA overestimates the actual transition temperature due
to the absence of Gaussian fluctuations, as discussed above.

\begin{figure}
\centering \includegraphics[width=0.8\columnwidth]{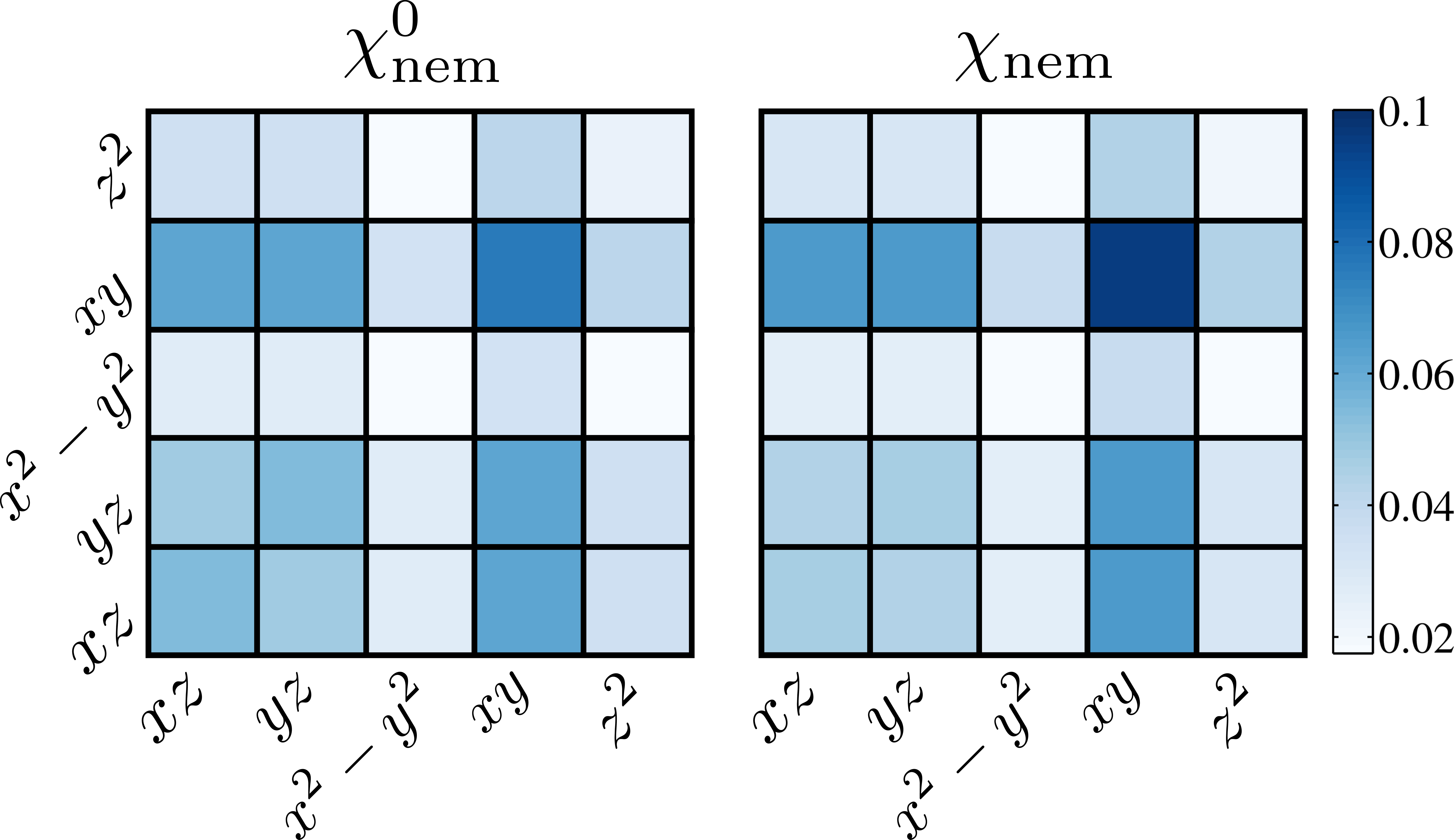}
\protect\protect\protect\protect\protect\protect\caption{\label{fig:nematic_and_magnetic_orbital_resolution} (Color online)
Color plot of the normalized elements of the eigen-matrix $\Phi_{\eta\mu}^{(n)}$
corresponding to the leading eigenvalue of the bare (left) and of
the full (right) nematic susceptibilities. The dominant contributions
arise from the $d_{xz}$, $d_{yz}$, and $d_{xy}$ orbital, with the
$d_{xy}$ being the most important for nematicity.}
\end{figure}

While the largest eigenvalue $\lambda^{(n)}$ determines $T_{\mathrm{nem}}$,
the structure of the corresponding $5\times5$ eigen-matrix $\Phi_{\eta\mu}^{(n)}$
reveals the orbital-resolved nematic order parameter driving the transition,
since $\Phi_{\eta\mu}^{(n)}\propto\left\langle \mbf{M}_{\eta}^{X}\mbf{M}_{\mu}^{X}\right\rangle -\left\langle \mbf{M}_{\eta}^{Y}\mbf{M}_{\mu}^{Y}\right\rangle $.
In Fig.~\ref{fig:nematic_and_magnetic_orbital_resolution} we plot
the normalized elements of the leading eigen-matrix $\Phi_{\eta\mu}^{(n)}$
for both the full and the bare nematic susceptibility -- which, as
shown above, contains information only about the magnetic instability.
In both cases, the dominant processes involve the $d_{xz}$, $d_{yz}$,
and $d_{xy}$ orbitals.

There is however one important difference: the relative weight of
the $d_{xy}$ orbital is larger for $\chi_{\text{nem}}^{\rho\nu\eta\mu}$
than for $\chi_{\text{nem},0}^{\rho\nu\eta\mu}$, i.e. while the three
orbitals seem to contribute equally to drive the magnetic instability,
the $d_{xy}$ orbital plays a more important role in driving the nematic
instability. We interpret this in terms of the nesting properties
of the orbital content of the Fermi surface in Fig.~\ref{fig:fermi_surface}:
while the $d_{xy}$ hole-pocket at $\left(\pi,\pi\right)$ can form
a single-\textbf{Q} SDW by combining with either the $X$ or $Y$
electron-pockets, since both have $d_{xy}$ spectral weight, the two
$d_{xz}$/$d_{yz}$ hole-pockets at $\left(0,0\right)$ can form a
double-\textbf{Q} SDW by combining with both the $X$ and $Y$ pockets,
since they have $d_{yz}$ and $d_{xz}$ spectral weight, respectively.

\begin{figure}
\centering \includegraphics[width=\columnwidth]{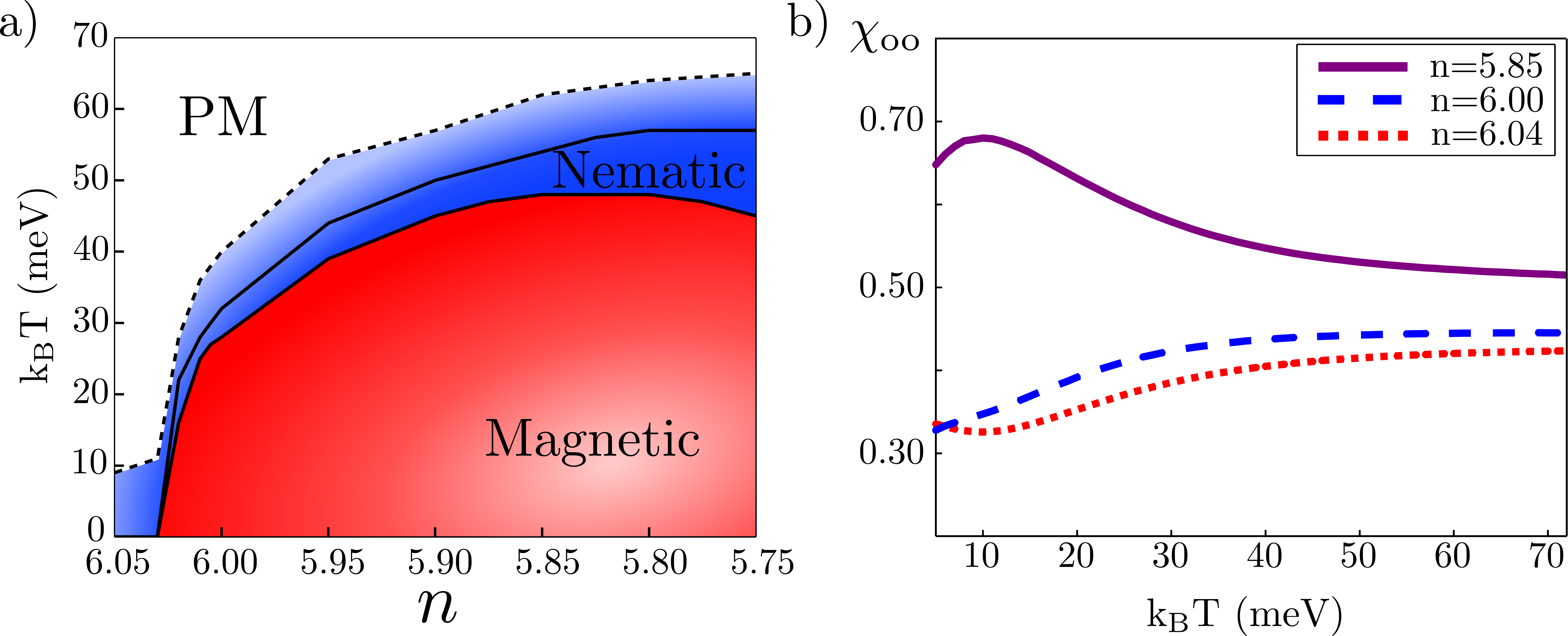}
\protect\protect\protect\protect\protect\protect\caption{\label{fig:phase_diagram} (Color online) \textbf{(a)} Occupation
number-temperature $(n,T)$ phase diagram for the bare magnetic and
nematic phase transitions, evidencing the narrow region displaying
nematic-paramagnetic order. The solid $T_{\mathrm{nem}}$ line takes
into account only the contribution from low-energy ($\Omega_{n}=0$)
magnetic fluctuations, whereas the dashed line includes contributions
from higher energies ($\Omega<\Omega_{c}=1$ eV). For $n<5.75$, an
incommensurate magnetic order appears. \textbf{(b)} Ferro-orbital
order susceptibility $\chi_{\mathrm{oo}}$ a function of $T$ for
various values of the occupation number $n$. In contrast to the nematic
susceptibility shown in Fig.~\ref{fig:nematic_and_magnetic_instabilities},
$\chi_{\mathrm{oo}}$ is nearly featureless and $T$-independent at
low energies.}
\end{figure}

Having analyzed the $n=6$ case, we present in Fig.~\ref{fig:phase_diagram}(a)
the complete $(n,T)$ phase diagram for the magnetic and nematic transitions.
We restrict our analysis to $n>5.75$, since below this value we find
incommensurate magnetic order. Accounting for the nematic transition
in this regime requires changes in the formalism beyond the scope
of this work. Note that, in contrast to experiments, $T_{\mathrm{mag}}$
is not peaked at $n=6$. This is likely due to the absence of disorder
effects introduced by doping, which are known to suppress $T_{\mathrm{mag}}$
\cite{Fernandes_Vavilov_12,Dagotto15}. Most importantly, across the
entire phase diagram the nematic transition line tracks closely the
magnetic transition line, in agreement with the phase diagrams of
the iron pnictides.

An important issue in obtaining this phase diagram is that, as shown
in Eq.~(\ref{eq:chi_bare_nem}), the computation of the nematic susceptibility
requires summing the magnetic fluctuations not only over the entire
Brillouin zone, but also over energy (i.e. over Matsubara frequencies).
Although the propagator $\chi_{i}^{\mu\nu}\left(\mathbf{q},\Omega_{n}\right)$
is strongly peaked at $\Omega_{n}=0$ (see Appendix \ref{app:freq_dep_mag_susc}), within
RPA it saturates to a finite value for large energies {[}see Eq. (\ref{eq:mag_susc}){]},
requiring a frequency cutoff $\Omega_{c}$. Near a finite-$T$ magnetic
transition, due to the very sharp peak in $\chi_{i}^{\mu\nu}\left(\mathbf{Q}_{X/Y},\Omega_{n}\right)$,
it is reasonable to take only the $\Omega_{n}=0$ contribution --
the low-energy magnetic fluctuations -- resulting in the solid line
of Fig.~\ref{fig:phase_diagram}. However, near the region where
$T_{\mathrm{mag}}\rightarrow0$, ignoring the high-energy magnetic
fluctuations ($\Omega_{n}\neq0$) is not justified. To address this
problem, we introduce a cutoff $\Omega_{c}=1$ eV, at which the propagator
reaches values close to its saturation value, as shown in Appendix \ref{app:freq_dep_mag_susc}.
The corresponding nematic transition line is shown as a dashed line
in Fig.~\ref{fig:phase_diagram}. Near the regime where the magnetic
transition takes place at finite $T$, the only effect of the cutoff
is to increase the nematic transition temperature, as expected. However,
near the regime where $T_{\mathrm{mag}}\rightarrow0$, the nematic
transition is stabilized even in the absence of long-range magnetic
order. Although the precise value of $T_{\mathrm{nem}}$ depends on
the cutoff value, the main result is that higher-energy magnetic fluctuations
are essential to promote nematic order without magnetic order. In
this regard, it is interesting to note that, in FeSe, the only parent
material in which nematic order is observed in the absence of magnetic
order, NMR measurements find no evidence for low-energy magnetic fluctuations
\cite{Buchner_FeSe,Meingast_FeSe}, whereas neutron scattering reports
sizable fluctuations at modest energy values \cite{INS_FeSe_1,INS_FeSe_2}.

A remaining question is whether or not the spin-driven nematic instability
is the leading instability of the system. In particular, an ongoing
debate \cite{fernandes02,kruger01,devereaux01,phillips01,ku01,Kontani12}
concerning iron-based materials is whether ferro-orbital order, signaled
by an unequal occupation of the $d_{xz}$ and $d_{yz}$ orbitals,
$\Delta n\equiv n_{xz}-n_{yz}\neq0$, could drive the nematic transition,
instead of the spin-driven mechanism explored above. To investigate
this issue, we calculate the $\mbf{q}=0$ static component of the
RPA orbital order susceptibility, $\chi_{\mathrm{oo}}(\mbf{q})=\left\langle \Delta n(\mbf{q})\Delta n(-\mbf{q})\right\rangle $
for the multi-orbital Hubbard model~\cite{Kontani15}, of which a brief derivation is included in Appendix \ref{app:ferro_orb_order_susc}.
As shown in Fig.~\ref{fig:phase_diagram}(b), our results reveal
a nearly $T$-independent $\chi_{\mathrm{oo}}$ for the doping range
and interactions investigated. This is not unexpected, since for reasonable
values of $U$ and $J$, there is no attraction in the RPA charge
channel. Therefore, within RPA, ferro-orbital order is unable to drive
the nematic instability. Of course, once the coupling to magnetic
fluctuations is included, which requires going beyond RPA, $\chi_{\mathrm{oo}}$
will diverge at the same $T$ as $\chi_{\mathrm{nem}}$ \cite{Nevidomskyy15,Kontani15,fanfarillo01}.
In this regard, by effectively decoupling these two channels, RPA
provides an interesting route to investigate which instability is
the leading one -- at least for weak or moderate interactions.

In summary, we developed an appropriate extension of the RPA approach
to obtain the orbital-resolved spin-driven nematic susceptibility
of an arbitrary multi-orbital Hubbard model. Application to the case
of iron-based superconductors reveals that the leading instability
of the system is an interaction-driven nematic phase. The $d_{xy}$
orbital plays a leading role in promoting the nematic instability,
and higher-energy magnetic fluctuations are essential to stabilize
nematic order in the absence of long-range magnetic order. Comparison
with other RPA susceptibilities reveals that the nematic and magnetic
transitions follow each other closely, and that the ferro-orbital
susceptibility does not diverge on its own. More generally, our formalism
can also be combined with first-principle approaches to search for
other materials that may display electronic nematicity. Furthermore,
because interactions appear only in the determination of the magnetic
propagator, Eq. (\ref{eq:mag_susc}), this formalism can be combined
with other approaches that specifically include moderate electronic
interactions, such as DFT+U or LDA+DMFT \cite{Medici13,park01}.

\begin{acknowledgments}
The authors are grateful to A. V. Chubukov, L. Fanfarillo, M. Sch{\"u}tt,
P. Orth, and B. Valenzuela for discussions. MHC and BMA acknowledge
financial support from a Lunbeckfond fellowship (grant A9318). RMF
and JK are supported by the U.S. Department of Energy, Office of Science,
Basic Energy Sciences, under award number DE-SC0012336. \end{acknowledgments}

\appendix

\section{Fourth order coefficients}\label{app:fourth_order_coeff}

To derive the form of the free energy given in Eqs. (2) and (4) in
the main text, we perform a Hubbard-Stratonovich (HS) decoupling thereby
obtaining the electron-mediated interactions between the magnetic
order parameters. Formally the HS decoupling relies on inserting unity
in the partition function, where unity, in the present case, is given
by 
\begin{eqnarray}
\mathds{1} &=& \int\mathcal{D}[\mbf{M}_{\mu\nu}^{X},\mbf{M}_{\mu\nu}^{Y}] \nonumber \\ && \exp\bigg[-\int_{q}\big(\mbf{M}_{\mu\nu}^{X}(q)\left(U^{-1}\right)_{\rho\lambda}^{\mu\nu}\mbf{M}_{\rho\lambda}^{X}(-q) \nonumber \\ && \qquad \qquad \ + \mbf{M}_{\mu\nu}^{Y}(q)\left(U^{-1}\right)_{\rho\lambda}^{\mu\nu}\mbf{M}_{\rho\lambda}^{Y}(-q)\big)\bigg]\,,
\end{eqnarray}
and $\int\mathcal{D}[\mbf{M}_{\mu\nu}^{X},\mbf{M}_{\mu\nu}^{Y}]$
is chosen such that the path-integral evaluates to unity and $q=(\mbf{q},\Omega_{n})$
($\Omega_{n}$ being a bosonic Matsubara frequency). The electrons
are then integrated out resulting in an effective action for the magnetic
order 
\begin{eqnarray}
\mathcal{S}_{\text{eff}}[\mbf{M}_{\mu\nu}^{X},\mbf{M}_{\mu\nu}^{Y}] &=& \sum_{i}\int_{q}\mbf{M}_{\mu\nu}^{i}(q)\left(U^{-1}\right)_{\rho\lambda}^{\mu\nu}\mbf{M}_{\rho\lambda}^{i}(-q) \nonumber \\ &-& \text{Tr}\ln\left[\mathbb{G}_{\mu\nu}^{0}(k)^{-1}-\mathcal{V}_{\mu\nu}(q)\right]\,,
\end{eqnarray}
where $i=X,Y$, $\mu$ and $\nu$ are orbital indices, $k=(\mbf{k},\omega_{n})$,
$\omega_{n}=\left(2n+1\right)\pi T$ is the fermionic Matsubara frequency,
and the trace is over all external indices (the spin indices have
been suppressed, the Green's function is diagonal in spin). $\mathbb{G}_{\mu\nu}^{0}(k)$
is the matrix Green's function, obtained from the first term in Eq.
(1) of the main text, and $\mathcal{V}$ originates from the coupling
between the magnetic order parameters and the electrons, the second
term. In the basis 
\begin{eqnarray}
\Psi(\mbf{k})=\begin{pmatrix} \psi(\mbf{k}) \\ \psi(\mbf{k+Q}_{X}) \\ \psi(\mbf{k+Q}_{Y}) \\ \psi(\mbf{k}+\mbf{Q}_{X}+\mbf{Q}_{Y})
\end{pmatrix}
\end{eqnarray}
these are given by the matrices
\begin{widetext}
\begin{eqnarray}
\mathbb{G}_{\mu\nu}^{0}(k) & = & \begin{pmatrix}\mathcal{G}_{\mu\nu}^{0}(k+q) & 0 & 0 & 0\\
0 & \mathcal{G}_{\mu\nu}^{0}(k+q+\mbf{Q}_{X}) & 0 & 0\\
0 & 0 & \mathcal{G}_{\mu\nu}^{0}(k+q+\mbf{Q}_{Y}) & 0\\
0 & 0 & 0 & \mathcal{G}_{\mu\nu}^{0}(k+q+\mbf{Q}_{X}+\mbf{Q}_{Y})
\end{pmatrix}\\
\mathcal{V}_{\mu\nu}(q) & = & \begin{pmatrix}0 & -\frac{1}{2}\mbf{M}_{\mu\nu}^{X}(q)\cdot\boldsymbol{\sigma}_{\alpha\beta} & -\frac{1}{2}\mbf{M}_{\mu\nu}^{Y}(q)\cdot\boldsymbol{\sigma}_{\alpha\beta} & 0\\
-\frac{1}{2}\mbf{M}_{\mu\nu}^{X}(q)\cdot\boldsymbol{\sigma}_{\alpha\beta} & 0 & 0 & -\frac{1}{2}\mbf{M}_{\mu\nu}^{Y}(q)\cdot\boldsymbol{\sigma}_{\alpha\beta}\\
-\frac{1}{2}\mbf{M}_{\mu\nu}^{Y}(q)\cdot\boldsymbol{\sigma}_{\alpha\beta} & 0 & 0 & -\frac{1}{2}\mbf{M}_{\mu\nu}^{X}(q)\cdot\boldsymbol{\sigma}_{\alpha\beta}\\
0 & -\frac{1}{2}\mbf{M}_{\mu\nu}^{Y}(q)\cdot\boldsymbol{\sigma}_{\alpha\beta} & -\frac{1}{2}\mbf{M}_{\mu\nu}^{X}(q)\cdot\boldsymbol{\sigma}_{\alpha\beta} & 0
\end{pmatrix}\,,
\end{eqnarray}
\end{widetext}
where each element of the matrices should be understood as an $N_{\text{orb}}\times N_{\text{orb}}$
matrix in orbital space, with the Green function being 
\begin{eqnarray}
\mathcal{G}_{\mu\nu}^{0}(k)=\sum_{m}\frac{\langle\mu|m\rangle\langle m|\nu\rangle}{i\omega_{n}-\xi^{m}(\mbf{k})}\,,
\end{eqnarray}
where $m$ refers to band basis and $\mu,\nu$ refer to orbital basis.
Expanding the trace-log to fourth order in the magnetic order parameters
and applying the Pauli matrix identity 
\begin{eqnarray}
\sigma_{\alpha\beta}^{i}\sigma_{\beta\delta}^{j}\sigma_{\delta\gamma}^{k}\sigma_{\gamma\alpha}^{l}=2\left(\delta^{ij}\delta^{kl}-\delta^{ik}\delta^{jl}+\delta^{il}\delta^{jk}\right)
\end{eqnarray}
yields the magnetic free energy as written in Eqs. (2) and (4) of
the main text, with the fourth order coefficients
\begin{widetext}
\begin{eqnarray}
u^{\rho\nu\eta\mu} & = & \frac{1}{16}\sum_{k}\Big(2\mathcal{G}^{\mu\rho}\mathcal{G}_{X}^{\rho\nu}\mathcal{G}^{\nu\eta}\mathcal{G}_{X}^{\eta\mu}-\mathcal{G}^{\mu\rho}\mathcal{G}_{X}^{\rho\eta}\mathcal{G}^{\eta\nu}\mathcal{G}_{X}^{\nu\mu}+\mathcal{G}^{\mu\rho}\mathcal{G}_{X}^{\rho\nu}\mathcal{G}^{\nu\eta}\mathcal{G}_{Y}^{\eta\mu}\nonumber \\
 & + & \mathcal{G}^{\nu\rho}\mathcal{G}_{X}^{\rho\mu}\mathcal{G}_{X+Y}^{\mu\eta}\mathcal{G}_{X}^{\eta\nu}-\mathcal{G}^{\mu\rho}\mathcal{G}_{X}^{\rho\eta}\mathcal{G}_{X+Y}^{\eta\nu}\mathcal{G}_{Y}^{\nu\mu}\Big)+(X\leftrightarrow Y)\,,\\
g^{\rho\nu\eta\mu} & = & -\frac{1}{16}\sum_{k}\Big(2\mathcal{G}^{\mu\rho}\mathcal{G}_{X}^{\rho\nu}\mathcal{G}^{\nu\eta}\mathcal{G}_{X}^{\eta\mu}-\mathcal{G}^{\mu\rho}\mathcal{G}_{X}^{\rho\eta}\mathcal{G}^{\eta\nu}\mathcal{G}_{X}^{\nu\mu}-\mathcal{G}^{\mu\rho}\mathcal{G}_{X}^{\rho\nu}\mathcal{G}^{\nu\eta}\mathcal{G}_{Y}^{\eta\mu}\nonumber \\
 & - & \mathcal{G}^{\nu\rho}\mathcal{G}_{X}^{\rho\mu}\mathcal{G}_{X+Y}^{\mu\eta}\mathcal{G}_{X}^{\eta\nu}+\mathcal{G}^{\mu\rho}\mathcal{G}_{X}^{\rho\eta}\mathcal{G}_{X+Y}^{\eta\nu}\mathcal{G}_{Y}^{\nu\mu}\Big)+(X\leftrightarrow Y)\,,\\
w^{\rho\nu\eta\mu} & = & \frac{1}{16}\sum_{k}\Big(-2\mathcal{G}^{\mu\rho}\mathcal{G}_{X}^{\rho\eta}\mathcal{G}^{\eta\nu}\mathcal{G}_{Y}^{\nu\mu}+2\mathcal{G}^{\nu\rho}\mathcal{G}_{X}^{\rho\eta}\mathcal{G}^{\eta\mu}\mathcal{G}_{Y}^{\mu\nu}-2\mathcal{G}^{\eta\rho}\mathcal{G}_{X}^{\rho\mu}\mathcal{G}_{X+Y}^{\mu\nu}\mathcal{G}_{X}^{\nu\eta}+2\mathcal{G}^{\eta\rho}\mathcal{G}_{X}^{\rho\nu}\mathcal{G}_{X+Y}^{\nu\mu}\mathcal{G}_{X}^{\mu\eta}\nonumber \\
 & + & \mathcal{G}^{\rho\mu}\mathcal{G}_{Y}^{\mu\eta}\mathcal{G}_{X+Y}^{\eta\nu}\mathcal{G}_{X}^{\nu\rho}+\mathcal{G}^{\rho\nu}\mathcal{G}_{Y}^{\nu\eta}\mathcal{G}_{X+Y}^{\eta\mu}\mathcal{G}_{X}^{\mu\rho}+\mathcal{G}^{\mu\rho}\mathcal{G}_{X}^{\rho\nu}\mathcal{G}_{X+Y}^{\nu\eta}\mathcal{G}_{Y}^{\eta\mu}+\mathcal{G}^{\nu\rho}\mathcal{G}_{X}^{\rho\mu}\mathcal{G}_{X+Y}^{\mu\eta}\mathcal{G}_{Y}^{\eta\nu}\Big)\,,
\end{eqnarray}
\end{widetext}
where repeated orbital indices are not summed. Here all the Green
functions are implicit functions of $k$ and $\mathcal{G}_{j}^{\mu\nu}(k)=\mathcal{G}^{\mu\nu}(k+\mbf{Q}_{j})$
and $\sum_{k}=T/N_{\mbf{k}}\sum_{\mbf{k}}\sum_{\omega_{n}}$.

\section{Nematic susceptibility}\label{app:nem_susc}

Preparing for an additional HS-decoupling we introduce two bosonic fields $\psi_{\rho\nu}$ and $\phi_{\rho\nu}$ with the partition function
\begin{eqnarray}
	\mathcal{Z} = \int \mathcal{D}\phi\mathcal{D}\psi \exp \Big[ && \frac{1}{2}\left(u^{\rho\nu\eta\mu} \right)^{-1}\psi_{\rho\nu}\psi_{\eta\mu} \nonumber \\ && - \frac{1}{2}\left(g^{\rho\nu\eta\mu}\right)^{-1}\phi_{\rho\nu}\phi_{\eta\mu} \Big]\,,
\end{eqnarray}
with integration measures chosen appropriately such that $\mathcal{Z}=1$. By performing the shifts
\begin{eqnarray}
	\psi_{\rho\nu} & \rightarrow \psi_{\rho\nu} - u^{\rho\nu\eta\mu}\left(\mbf{M}_{\eta}^{X}\cdot\mbf{M}_{\mu}^{X} + \mbf{M}_{\eta}^{Y}\cdot\mbf{M}_{\mu}^{Y} \right)\,, \\ 
	\phi_{\rho\nu} & \rightarrow \psi_{\rho\nu} + g^{\rho\nu\eta\mu}\left(\mbf{M}_{\eta}^{X}\cdot\mbf{M}_{\mu}^{X} - \mbf{M}_{\eta}^{Y}\cdot\mbf{M}_{\mu}^{Y} \right)\,,
\end{eqnarray}
the terms quartic in $\mbf{M}$ cancel accordingly. Following the standard procedure we introduce a field ($h_{\rho\nu}$) conjugate to $\mbf{M}^{X}_{\rho}\cdot \mbf{M}^{X}_{\nu} - \mbf{M}^{Y}_{\rho}\cdot \mbf{M}^{Y}_{\nu}$ and define $\widetilde{\phi}_{\rho\nu}=\phi_{\rho\nu}+h_{\rho\nu}$. The resulting action is then
\begin{eqnarray}
	\mathcal{S}[\mbf{M}^{i}_{\mu},\psi_{\mu\nu},\phi_{\mu\nu}] &=& \sum_{q,i=X,Y} \left(r^{\mu\nu}_{i}(q) + \psi_{\mu\nu}\right) \mbf{M}^{i}_{\mu}\cdot\mbf{M}^{i}_{\nu} \nonumber \\ &-& \frac{1}{2}\left( u^{\rho\nu\eta\mu} \right)^{-1} \psi_{\rho\nu}\psi_{\eta\mu} \nonumber \\ &+& \frac{1}{2} \left( g^{\rho\nu\eta\mu} \right)^{-1} \left( \widetilde{\phi}_{\rho\nu} - h_{\eta\mu} \right) \left( \widetilde{\phi}_{\eta\mu} - h_{\eta\mu} \right) \nonumber \\
 &-& \widetilde{\phi}_{\rho\nu}\left( \mbf{M}^{X}_{\rho}\cdot \mbf{M}^{X}_{\nu} - \mbf{M}^{Y}_{\rho}\cdot\mbf{M}^{Y}_{\nu} \right)\,.
\end{eqnarray}
Here $r^{\mu\nu}_{i}(q)=(U^{\mu}_{\nu})^{-1} + \sum_{k}\mathcal{G}^{\nu\mu}(k)\mathcal{G}^{\mu\nu}_i(k+q)$ and $\mathcal{G}^{\mu\nu}_i(k) \equiv \mathcal{G}^{\mu\nu}(k+\mbf{Q}_{i})$. It is now straightforward to compute the nematic susceptibility:
\begin{eqnarray}
	\chi^{\rho\nu\eta\mu}_{\text{nem}} &=& \lim_{h\rightarrow 0} \left( \frac{\delta^{2} \ln \mathcal{Z}}{\delta h_{\rho\nu}\delta h_{\eta\mu}} \right) \nonumber \\
	&=& \left(g^{\rho\nu\iota\kappa}g^{\eta\mu\phi\lambda} \right)^{-1} \left\langle \phi_{\iota\kappa} \phi_{\phi\lambda} \right\rangle - \left( g^{\rho\nu\eta\mu} \right)^{-1}\,,
\end{eqnarray}
where we used the fact that $\left\langle \phi_{\rho\nu} \right\rangle =0$ as we are above the nematic instability. To continue we note that
\begin{eqnarray}
	\frac{\delta^2 F}{\delta \phi_{\rho\nu} \delta \phi_{\eta\mu}} = \left\langle \phi_{\rho\nu} \phi_{\eta\mu} \right\rangle^{-1}\,,
\end{eqnarray}
where the free energy is
\begin{eqnarray}
	F= -T\ln \mathcal{Z},
\end{eqnarray}
obtained by integrating out the magnetic degrees of freedom and taking the large $N$ limit. We find the effective action
\begin{eqnarray}
	\mathcal{S}_{\text{eff}}[\psi_{\mu\nu},\phi_{\mu\nu}] &=& \frac{1}{2}\left( g^{\rho\nu\eta\mu} \right)^{-1}\phi_{\rho\nu}\phi_{\eta\mu} \nonumber \\ &+& \frac{1}{2} \text{Tr} \ \ln \big[ \chi^{-1}_{\iota\kappa,Y}\chi^{-1}_{\kappa\lambda,X} - \phi_{\iota\kappa}\phi_{\kappa\lambda} \nonumber \\ &+& \chi^{-1}_{\iota\kappa,Y}\phi_{\kappa\lambda} - \phi_{\iota\kappa}\chi^{-1}_{\kappa\lambda,X} \big]\,,
\end{eqnarray}
where we have ignored the Gaussian fluctuations $\psi_{\rho\nu}$ and $(\chi^{\mu\nu}_{i}(q))^{-1}=r^{\mu\nu}_{i}(q)$. Finally
\begin{eqnarray}
	\left\langle \phi_{\rho\nu}\phi_{\eta\mu} \right\rangle^{-1} & = & \left( g^{\rho\nu\eta\mu} \right)^{-1} \nonumber \\ &-& \frac{1}{2}\sum_{q,i=X,Y} \chi_{\rho\mu,i}(q)\chi_{\nu\eta,i}(-q)
\end{eqnarray}
and after some manipulations we arrive at the expression given in the text for the nematic susceptibility.

\section{Frequency dependence of the magnetic susceptibility}\label{app:freq_dep_mag_susc}

\begin{figure}[t]
\centering \includegraphics[width=\columnwidth]{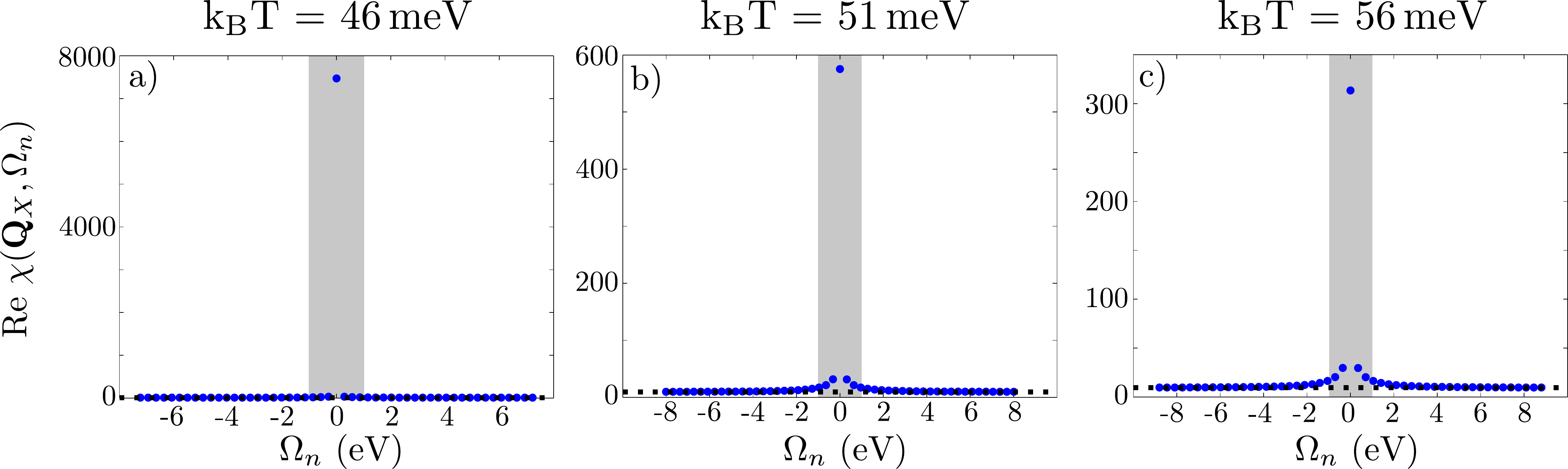}
\protect\protect\caption{\label{fig:frequency_dependence_n=00003D00003D590} Frequency dependence
of the magnetic propagator $\sum_{\mu\nu}\chi_{X}^{\mu\nu}(\mbf{Q}_{X},\Omega_{n})$
for $n=5.90$ at different temperatures. The parameters used are quoted
in the main text. The magnetic instability takes place at $k_{B}T=45$
meV. From (a) we see that the contribution to the bare nematic susceptibility
comes mostly from the zero frequency part of the magnetic susceptibility.}
\end{figure}

In this section we illustrate the frequency dependence of the magnetic
propagator at various temperatures for representative filling factors
of the $(n,T)$ phase diagram (Fig. 4(a) of the main text). Because
the magnetic propagator peaks at $(\pi,0)/(0,\pi)$, we focus on $\mbf{Q}_{X}$.
For $n=5.90$ as we approach the instability (at $k_{\text{B}}T=45$
meV), the frequency dependence of the propagator $\sum_{\mu\nu}\chi_{X}^{\mu\nu}(\mbf{Q}_{X},\Omega_{n})$
has the form shown in Fig. \ref{fig:frequency_dependence_n=00003D00003D590},
where the bosonic Matsubara frequency is given by $\Omega_{n}=2\pi nT$.
The gray area denotes the region included in the cut-off $\Omega_{c}=1\,\text{eV}$,
and the dotted line indicates $\sum_{\mu\nu}\chi_{X}^{\mu\nu}(\mbf{Q}_{X},\Omega_{n}\rightarrow\infty)$.
The plots in Fig. \ref{fig:frequency_dependence_n=00003D00003D590}
justify the statement made in the main text that near a finite-temperature
magnetic transition, one can safely neglect the higher frequency contributions.

\begin{figure}[t]
\centering \includegraphics[width=\columnwidth]{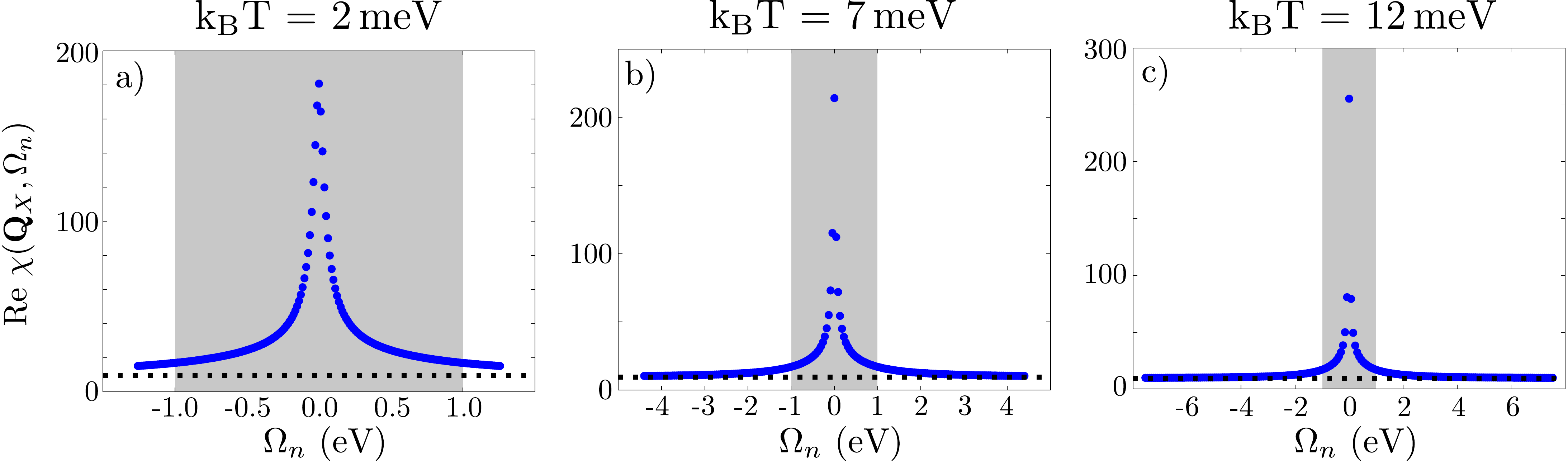}
\protect\protect\caption{\label{fig:frequency_dependence_n=00003D00003D604} Frequency dependence
of the magnetic propagator $\sum_{\mu\nu}\chi_{X}^{\mu\nu}(\mbf{Q}_{X},\Omega_{n})$
for $n=6.04$ at different temperatures. The parameters used are quoted
in the main text. As is evident in (a), the peak broadens as zero
temperature is approached. However, even at higher temperatures, shown
in (b) and (c), finite Matsubara frequencies provide considerable
contributions to the bare nematic susceptibility.}
\end{figure}

To illustrate the importance of including high frequency contributions
in the case where magnetic order is absent, in Fig. \ref{fig:frequency_dependence_n=00003D00003D604}
we also plot the frequency dependence of the magnetic propagator for
$n=6.04$. It is clear that the peak is broadened, implying that it
is no longer justified to ignore the contributions originating from
finite frequencies.

\section{Derivation of the ferro-orbital order susceptibility}\label{app:ferro_orb_order_susc}

Ferro-orbital order is characterized by the breaking of the degeneracy
between the $d_{xz}$ and $d_{yz}$ orbitals. In the itinerant framework
this is seen by an inequivalent occupation of the two orbitals, i.e.
$n_{xz}\neq n_{yz}$. Defining $\Delta n(q)\equiv n_{xz}(q)-n_{yz}(q)$
as in the main text, the ferro-orbital susceptibility is given by
$\langle\Delta n(q)\Delta n(-q)\rangle$. Using the definition of
$\Delta n(q)$, we find that this is nothing but a linear combination
of specific components of the charge susceptibility, $(\chi^{\mathrm{c}})_{\rho\lambda}^{\mu\nu}$.
In the standard RPA approach, the full expression is \cite{Kontani15,Nevidomskyy15}
\begin{eqnarray}
\chi_{\mathrm{oo}} &=& (\chi_{\text{RPA}}^{\mathrm{c}})_{xz,xz}^{xz,xz}+(\chi_{\text{RPA}}^{\mathrm{c}})_{yz,yz}^{yz,yz} \nonumber \\ &-& (\chi_{\text{RPA}}^{\mathrm{c}})_{xz,yz}^{xz,yz}-(\chi_{\text{RPA}}^{\mathrm{c}})_{yz,xz}^{yz,xz}\,,\label{eq:chi_oo}
\end{eqnarray}
where the RPA charge susceptibility is given by the usual expression
\cite{kemper01} 
\begin{eqnarray}
(\chi_{\text{RPA}}^{\mathrm{c}})_{\rho\lambda}^{\mu\nu}=\left(\left[1+\chi_{0}U_{c}\right]^{-1}\right)_{\rho\gamma}^{\mu\delta}(\chi_{0})_{\gamma\lambda}^{\delta\nu}\,,\label{eq:charge_rpa}
\end{eqnarray}
where $\chi_{0}$ is the standard particle-hole bubble 
\begin{eqnarray}
(\chi_{0}(q))_{\rho\lambda}^{\mu\nu} & = & -\sum_{k}\mathcal{G}^{\mu\nu}(k)\mathcal{G}^{\rho\lambda}(k+q)
\end{eqnarray}
and $U_{c}$ is the interaction matrix in the charge channel. The
latter differs from the interaction in the SDW channel and is given
by ($a\neq b$) 
\begin{eqnarray}
(U_{c})_{aa}^{aa} & = & U\,,\\
(U_{c})_{bb}^{aa} & = & 2U'-J=2U-5J\,,\\
(U_{c})_{ab}^{ab} & = & 2J-U'=4J-U\,,\\
(U_{c})_{ab}^{ba} & = & J'=J\,.
\end{eqnarray}
We note that, due to the implicit summation over repeated indices
in Eq. (\ref{eq:charge_rpa}), all orbitals contribute to the RPA
orbital order susceptibility. The static part of Eq. (\ref{eq:chi_oo})
at $\mbf{q}=0$ is the quantity plotted in Fig. 4(b) in the main text.

\end{document}